\documentclass[conference]{IEEEtran}

\usepackage[utf8]{inputenc}
\usepackage{graphicx,bm,color}
\usepackage[dvipsnames]{xcolor}
\usepackage[margin=0.95in]{geometry}
\usepackage{amsthm,amsmath,amssymb,mathtools,bm}
\usepackage{multirow}
\usepackage{comment}
\usepackage{wrapfig}
\usepackage{ragged2e}
\usepackage[normalem]{ulem}
\usepackage{hyperref}
\usepackage{cite}

\usepackage{enumitem}    %
\usepackage{ragged2e}    %
\usepackage{algorithm}   %
\usepackage{amsmath}     %

\DeclareMathSizes{10}{10}{7}{5}
\DeclareMathSizes{8}{8}{6}{5}
\DeclareMathSizes{7}{7}{5}{5}

\usepackage{caption}
\usepackage[noend]{algpseudocode}
\usepackage{algorithm}

\input{Qcircuit}

\newcommand{\sh}[1]{#1}

\newcommand{\ketbra}[2]{\ket{#1}\!\bra{#2}}

\newcommand{\reals}{{\mathbb R}}
\newcommand{\integers}{{\mathbb Z}}

\newtheorem{rem}{Remark}

\newcommand{\noq}{n}

\newcommand{\x}{\mathbf{x}}
\newcommand{\y}{\mathbf{y}}

 \theoremstyle{plain}
 
 \theoremstyle{plain}
 \theoremstyle{plain}
 
 \theoremstyle{plain}
 \theoremstyle{plain}
 
 \theoremstyle{plain}
 
 \theoremstyle{plain}
 
 \theoremstyle{remark}
 \newtheorem*{rem*}{Remark}
 \theoremstyle{plain}

\theoremstyle{plain}
 \newtheorem*{conj*}{Conjecture}
 \theoremstyle{plain}

\begin{document}

\title{Noise-Directed Adaptive Remapping for Integer Optimization: from qubits to (encoded) qudits}

\author{\IEEEauthorblockN{Stuart Hadfield~\IEEEmembership{Member,~IEEE}}
\IEEEauthorblockA{
\textit{USRA RIACS}\\
\textit{NASA Quantum AI Laboratory}\\
Moffett Field, CA, USA\\
shadfield@usra.edu}
\and
\IEEEauthorblockN{Filip B. Maciejewski}
\IEEEauthorblockA{
\textit{USRA RIACS}\\
Moffett Field, CA, USA\\
fmaciejewski@usra.edu}
\and
\IEEEauthorblockN{Davide Venturelli}
\IEEEauthorblockA{
\textit{USRA RIACS}\\
\textit{NASA Quantum AI Laboratory}\\
Moffett Field, CA, USA\\
dventurelli@usra.edu}
}

\date{\today}

\maketitle

\begin{abstract}
We extend Noise-Directed Adaptive Remapping (NDAR), 
a recently proposed heuristic meta-algorithm that leverages device noise as a computational resource, 
to optimization problems over discrete (integer) domains. While originally introduced for unconstrained binary optimization, the proposed generalization introduces additional gauge degrees of freedom at the logical level, such that the gauge transformation applied at each iteration is no longer unique, 
allowing tailoring to particular encodings or quantum hardware. 
We identify encoding-dependent requirements for NDAR
beyond binary domains: feasibility of the noise attractor, existence of compatible gauge transformations %
that preserve an efficiently implementable circuit family, and a %
systematic way to select the transform to apply at each step.  
We analyze these criteria for qudit-native and for binary, one-hot, %
and domain-wall qubit encodings, using the Max-$k$-colorable subgraph problem as a
running example.
We demonstrate that these encodings can
exhibit distinct advantages and tradeoffs when integrated
within the NDAR framework, particularly in how noise-induced dynamics interact with the solution landscape and
choice of encoding. Our results indicate that NDAR-guided
noise considerations provide a new criterion for comparing
device-level encoding choices for quantum optimization. Finally, we outline directions toward experimental realization
in superconducting qudit devices and further algorithmic
improvements.

\end{abstract}


\section{Introduction}
Noise-Directed Adaptive Remapping (NDAR) 
is a meta-algorithm 
recently proposed in \cite{maciejewski2024improving} and demonstrated to dramatically 
improve the performance of quantum optimization algorithms deployed on real-world quantum hardware. 
The %
theory was developed for unconstrained binary optimization problems and quantum ans\"atze related to the 
Quantum Approximate Optimization Algorithm (QAOA)~\cite{farhi2014quantum,hadfield2019quantum}, and a convincing experimental demonstration was performed on a Rigetti Computing quantum processing unit (QPU), showing significant improvement with NDAR over the use of the quantum ansatz alone~\cite[Fig. 2]{maciejewski2024improving}. %
Given the many constraints and limitations of near-term quantum hardware~\cite{awschalom2025challenges} and algorithms~\cite{abbas2024challenges}, NDAR proposed a radically new approach to dealing with certain types of hardware noise: 
rather than attempting to error-correct or mitigate its effects, NDAR seeks to \textit{exploit} noise as a potential resource leading to improved algorithm performance.

The main idea of NDAR is the following. Certain common %
types or components of noise  
cause the quantum computer to behave in a predictable way – the noise dynamics \sh{effectively} tries to steer %
the quantum computing output towards some preferred “attractor” state. 
For example, in systems or models where noise favors states of lower Hamming weight (excitation number), 
such as due to amplitude damping or %
dissipative effects, 
the attractor corresponds to the %
all-zero state. 
NDAR utilizes knowledge of the attractor state 
in combination with data drawn from noisy quantum measurements  
to implement mathematical operations (“gauge transforms”) that dynamically adjust %
how the problem is logically mapped to the quantum device. 
It further exploits the important observation that %
many quantum optimization algorithms require repeated preparation and measurement, which yields a large pool of candidate solutions that can be leveraged.  
Specifically, %
at each iteration NDAR 
transforms the cost Hamiltonian so as to 
align the noise attractor with %
the best candidate solution found. %

\begin{table*}[h]
\small
\centering
\begin{tabular}{|c || c || c | c | c | c |} 
 \hline
 Encoding & Qudit & Binary / Gray & One-Hot & 01-Hot & Domain-Wall \\ [0.5ex] 
 \hline
 Qudits/Qubits per variable & 1& $\lceil \log_2 d \rceil $ & $d$ & $d-1$ & $d-1$ \\ 
 \hline
 Probability $q$ of valid string & $q=1$ & %
 $q=1$ 
 & $q \leq 1$ & $q \leq 1$  & $q \leq 1$ \\
 \hline
Additional hard constraints & No & No & Yes & Yes & Yes \\
 \hline
  $\ket{00\dots 0}$ attractor feasible & Yes  & Yes & No & Yes & Yes  \\
 \hline
 Cost Hamiltonian / Phase Operator & %
 2-local & 2$\lceil \log_2 d \rceil$-local & 2-local & $2(d-1)$-local & 2-local \\
 \hline
 Mixing Operator & 1-local & 1-local & 2-local & $(d-1)$-local & 2-local \\
 \hline 
\end{tabular}
\caption{Comparison of $d$-level encodings on qudits and qubits for problems over $\mathbb{Z}_d^n$ such as  %
Max-$d$-colorable 
subgraph. 
Native qudit encodings %
exhibit a 
\lq\lq \sh{best-of-class}\rq\rq\
combination of %
advantageous properties over qubit ones. %
\sh{The binary column is immediate when $d$ is a power of $2$, or assuming a many-to-one encoding otherwise.}
}

\label{tab:summary}
\end{table*}

Building %
on the results of \cite{maciejewski2024improving}, we explore how NDAR may be extended to applications beyond unconstrained binary optimization, and to more general quantum devices, including those based on $d$-level quantum systems (qudits). %
Indeed, a number of emerging quantum technologies are %
moving beyond strictly two-level quantum systems, utilizing multiple quantum levels within each quantum subsystem, i.e., \textit{qudit}-based quantum processors \cite{wang2020qudits,su2021construction,Blok2021QutritScrambling,chi2022programmable,Ringbauer2022UniversalQudit,roy2023two,denys2023,nguyen2024empowering,bornman2025benchmarking,kim2025ultracoherent,venturelli2025near}.
Here we focus on problems over discrete domains such as $\mathbb{Z}_d^n$, which can be mapped directly to such systems or to qubits via different %
encodings, %
including both binary and unary variants~\cite{hadfield2019quantum,chancellor2019domain}. 
To this end we generalize NDAR beyond prior work to explicitly handle certain classes of \textit{hard constraints} which may be given as problem input, or may arise directly from the choice of encoding. 
\sh{Numerous practically important optimization tasks, %
for example scheduling or routing problems, are naturally formulated using variables taking $d>2$ possible values~\cite{hadfield2019quantum,sawaya2022encoding}.} %
For the prototypical %
integer domain problems we consider, we show that %
direct qudit implementations of NDAR present several advantages over comparable qubit-based encodings; %
some of the key differences  
are summarized in Table~\ref{tab:summary}, with details given in  Secs.~\ref{sec:ndarQudits} and~\ref{sec:ndarencoded}. 
We leverage our results to give an outlook for experimental realizations in the near-term and beyond in Sec.~\ref{sec:discussion}.

\section{Preliminaries}
We first %
overview some key technical components %
of quantum optimization, extensions to qudits, and NDAR. %
For simplicity we present a single overarching version of NDAR in Algorithm 1 below that %
subsumes prior work as special cases.

\subsection{Combinatorial Optimization}
Consider optimization problems over strings %
$y\in\integers_k^n$, where we are given a cost function %
\begin{equation}
    c:\integers_k^n\rightarrow \reals
\end{equation}
that we seek to maximize (or minimize). 
The case $k=2$ corresponds to binary optimization. The case $k>2$ captures the case of each variable drawn from one of $k$ possible assignments (which are identified with $k$ integers). For simplicity we assume each variable is drawn from the same domain (same $k$), whereas in general they may be different, and that $k$ is finite.  

Additionally, we may also be given a set $\{a_j\}$ of hard constraints 
$a_j:\integers_k^n\rightarrow \reals$ 
with the requirement that 
\begin{equation} \label{eq:constraints}
    \forall j\,\,\,a_j(y)=0
\end{equation}
must be satisfied for any valid candidate solution $y$; such strings~$y$ are called \textit{feasible}. 
While for unconstrained optimization this set is initially empty, the choice of problem encoding can nevertheless introduce hard constraints; the latter is the type of problem constraints we primarily consider in this paper.
Hence %
we will sometimes encounter \textit{invalid} strings in the encoded domain that %
do not correspond to %
variable assignments. %
For example, the $00\dots0$ string  is invalid in a one-hot encoding of an integer variable. %
Invalid strings can often be efficiently \lq\lq corrected\rq\rq\ to obtain a valid string using a fixed heuristic, though often lacking closeness guarantees. For example, for one-hot encodings, a measured string of Hamming weight greater than one could flip bits to zero until valid, say randomly, or based on the structure of the problem such as which correction is most advantageous. %

For quantum optimization algorithms on qubits, such as those based on QAOA and its generalizations, variable values are typically mapped to qubit basis states~$\ket{0},\ket{1}$. 
The cost function is then  mapped~\cite{lucas2014ising,hadfield2021representation} to a diagonal cost Hamiltonian to be maximized (or minimized):
\begin{align}\label{eq:cost_hamiltonian}
    H = h_0 I +\sum_{i} h_{i} Z_i + \sum_{i<j} h_{ij} Z_iZ_j +\dots
\end{align}
where $Z_i$ is the Pauli $Z$ operator acting on qubit $i$,  $h_i$ is a local field magnitude, and $h_{ij} \in \mathbb{R}$ is an interaction strength between qubits $i,j$. 
When hard constraints are present, each $a_j$ (or $a_j^2$) of \eqref{eq:constraints} can be similarly mapped to diagonal Hamiltonians $A_j$ that together act to penalize infeasible or invalid states. 

We give a qudit generalization of %
\eqref{eq:cost_hamiltonian} in Sec.~\ref{sec:QAOAqudits}. %

\subsubsection{Bitflip transforms} 
For qubits, consider the unitary bitflip (spin-reversal) operators $P_{\y} = \bigotimes_{i=0}^{\noq-1}X_{i}^{y_i}$ that act to flip the $|0\rangle$, $|1\rangle$ basis states for each bit indicated by the bitstring $y\in\{0,1\}^\noq$. 
For bitflip transforms we have the property $P_{\y}^\dagger=P_{\y}$, which will not hold for the more general gauge transformations we consider shortly. 

Applying this change-of-basis to $H$ 
is equivalent to changing the 
signs of the coefficients $h_i$ and $h_{ij}$ as 
$H \rightarrow H^{\y}:=P_{\y}^\dagger HP_{\y}$ with 
\begin{align}\label{eq:gauge_transformation_local}
    H^{\y} 
    = %
    P_{\y} HP_{\y} \; &= h_0 I
    +\sum_{i} \left(-1\right)^{y_i}h_{i} Z_i \nonumber \\
    &+ \sum_{i<j} \left(-1\right)^{y_i+y_j}h_{ij} Z_iZ_j +\dots,
\end{align}
and similarly for any higher order terms. 
This gauge transformation preserves cost Hamiltonian eigenvalues, with eigenvectors (candidate problem solutions) permuted~under $P_{\y}$. 
In particular, under this transformation the $|0\dots0\rangle$ state is mapped to~$|y_0\dots y_{n-1}\rangle$. 
Any constraint Hamiltonians $A_j$ are also diagonal and transform in the same way as~$H$. 

For QAOA, both the standard initial state and mixing operator are invariant under $P_{\y}$, and the phase operator changes only in sign of the coefficients that correspond to the changes to $H$, which will compile to %
the same circuit up to negation of the rotation angles of a subset of quantum gates.

\begin{figure}
    \centering
    \includegraphics[width=0.9\linewidth]{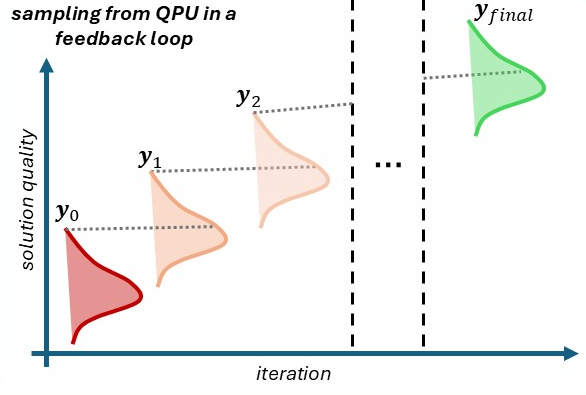}
    \caption{Schematic illustration of NDAR: the distribution of solution costs obtained from the quantum device is iteratively improved by successive problem gauge transforms which seek to align the noise attractor with high-quality solutions; see e.g.~\cite[Fig. 1]{maciejewski2024improving} for comparable results obtained from real-world quantum hardware.  }
    \label{fig:fig1small}
\end{figure}

\subsection{Noise-Directed Adaptive Remapping (NDAR)} \label{sec:ndar}

NDAR is a recently proposed meta-algorithm~\cite{maciejewski2024improving} that can significantly improve quantum optimization performance on device implementations by exploiting rather than mitigating the effects of hardware noise. 
Here we recap the main ideas and requirements towards generalization. 
NDAR works by identifying a classical ``attractor state" of the noise naturally present in the quantum processor, and then dynamically remapping how the problem is encoded on the quantum device to use the noise as an active local search heuristic. %
At each iteration, a gauge transform (such as the bitflip transform) is determined via data sampled from the quantum runs, adaptively selected to improve the cost value corresponding to the attractor. %

\begin{algorithm}
\caption{%
\centering NDAR for qudits or qubits (greedy remapping)}\label{alg:NDAR_main}
\scriptsize

\RaggedRight\textbf{Input}: ($\ket{00\dots 0}$ is the assumed feasible attractor state)
\begin{itemize}[leftmargin=*, itemsep=2pt, topsep=2pt]
    \item $H_{(0)}$: cost Hamiltonian to be maximized. (Minimization case is similar).
    \item $\{A_i\}$: initial hard constraints that must be satisfied.
    \item \texttt{optimizer}: a quantum subroutine that performs a stochastic optimization of $H_{(l)}$ (gauge-transformed Hamiltonian in iteration step $l+1$).
    \item gauge transformation selection rule $\x\rightarrow P_\x$
    \item $M$: total number of samples in each optimization step.
    \item \texttt{TERMINATION}: a rule specifying when to halt.
\end{itemize}
\vspace{2pt}

\RaggedRight\textbf{Pseudocode for iteration $j$}:
\begin{enumerate}[leftmargin=*, itemsep=2pt, topsep=2pt]
    \item Run \texttt{optimizer} for $H_{(j-1)}$, $\{A_i\}$. Result: $M$ strings $\{\x^i_{(j)}\}_{i=1}^{M}$.
    \item Discard or correct any infeasible strings, yielding $M' \leq M$ solutions.
    \item Calculate costs $\{E_{(j)}^{i} = \langle\x^i_{(j)}| H_{(j-1)} |\x^i_{(j)}\rangle\}_{i=1}^{M'}$.
    \item Select the string $\x^{\ast}$ such that  %
    $E_{(j)}^{\ast} = \max_{i} E_{(j)}^i$ (or $=\min_{i} E_{(j)}^i$).\\
    \vspace{2pt}
    \textbf{if} \texttt{TERMINATION}: \textbf{return} highest (lowest) energy solution found.
    \item Apply gauge transform $P_{\x^{\ast}}$ to $H_{(j-1)}$ to obtain $H_{(j)} \coloneqq H_{(j-1)}^{\x^{\ast}}=P_{\x^{\ast}}^\dagger H_{(j-1)} P_{\x^{\ast}}$ (cf.~Eq.~\eqref{eq:gauge_transformation_local}). This ensures that the transformed Hamiltonian satisfies $\bra{00\dots 0} H_{(j)} \ket{00\dots 0} = E^{\ast}_{(j)}$.
    \item Apply $P_{\x^{\ast}}$ to each constraint: $A_i \leftarrow P_{\x^{\ast}}^\dagger A_i P_{\x^{\ast}}$.
    \item Increment $j$ and proceed to Step 1.
\end{enumerate}
\end{algorithm}

Prior work considered NDAR for qubits in the setting of unconstrained optimization. 
Abstracted toward generalization, NDAR rests on four key ingredients:
\begin{enumerate}
    \item \textbf{Attractor:} a noise model or device behavior with a known classical \emph{attractor state} (e.g., $\ket{00\dots 0}$).
    
    The attractor may follow from the noise model or be inferred empirically. For instance, any noise channel that suppresses Hamming weight makes $\ket{00\dots 0}$ the attractor, regardless of how often it is actually sampled.
    
    \item \textbf{Feasibility:} the attractor must encode a \emph{feasible solution} at every iteration.
    
    This is automatic for unconstrained binary optimization, but is not guaranteed once hard constraints or non-trivial encodings enter the picture.
    Here we extend NDAR to families of constraints $\{A_i\}$ including those whose feasible sets form a transitive orbit under an efficiently constructible group action. This class includes, but is not limited to, common encodings of integer domains. %
    For example, in one-hot encodings all feasible codewords lie in a single orbit of an easily implemented permutation. In contrast, for the Maximum Independent Set problem the feasible set is not generally a transitive orbit under simple permutations, and as a result we cannot construct a set of gauge transformations satisfying all of our criteria in a straightforward way. %

    \item \textbf{Gauge %
    selection:} 
each accepted measurement outcome must \emph{uniquely determine}, through a specified selection rule, a gauge transformation \(x\mapsto P_x\) that maps the attractor to the target string.
    
    For binary optimization, the bitflip (automorphism) group acts \emph{simply transitively} on the set of bitstrings, so the best sampled solution unambiguously specifies the next gauge. Beyond binary, this correspondence breaks and the mapping must be specified.
    
    To choose the target string at each iteration here we follow the \textit{greedy choice rule}~\cite{maciejewski2024improving} of mapping to the best solution found at the current iteration. 
    
    \item \textbf{Practical circuit remappings:} the derived quantum circuit must transform in a way under the gauge that preserves sufficient structure of the ansatz so as to not introduce unscalable overheads in the iterative procedure. This is ultimately an operational constraint that will depend on implementation details (such as compilation efficiencies) as well as hardware specifications and noise.
    
    Bitflip gauges %
    leave QAOA circuits identical up to %
    signs of (single-qubit) rotations: hence the noise profile is %
    preserved across iterations. More general transformations may introduce new terms or dramatically reshape the circuit, breaking this property.

\end{enumerate}

We show the steps of the simplest version of NDAR in Algorithm 1, presented generally to account for the extensions to qudits and encoded problems of Sec. \ref{sec:QAOAqudits}. Our Steps 2 and 6 are new, to explicitly deal with infeasible strings in cases of problems with hard constraints. %
Note that Algorithm 1 assumes the attractor state to be $\ket{00\dots 0}$, encoding a feasible state. %

\paragraph*{Remarks on hard constraints} 
Different approaches exist 
for extending quantum optimization algorithms to 
problems with hard constraints, with different tradeoffs.
A common approach is to incorporate constraints directly into %
a modified cost Hamiltonian %
$H'=H-\lambda \sum_j A_j^2$ that contains 
sufficiently large 
penalty terms %
as controlled by the %
\sh{parameter~$\lambda>0$ (or $\lambda <0$ for minimization).}
This approach still yields infeasible solutions with nonnegligible probability. 
Alternatively, one can avoid this overhead by instead incorporating constraints into the design of mixer operators and initial state such that the algorithm is restricted to the subspace of feasible states~\cite{hadfield2019quantum,sawaya2022encoding,leipold2024imposing}. 
Our assumption that the attractor state is \sh{initially} feasible, together with the property that our modified algorithm %
only selects gauge transforms \sh{that map it} to other feasible states, %
implies that our generalization of NDAR is compatible with both approaches.

\subsection{Quantum Optimization with Qudits} \label{sec:QAOAqudits}

Abstractly, the domain $\integers_d^n$ naturally corresponds to the state space of $n$ %
$d$-level qudits, in the same way as for binary domains and qubits. Hence, irrespective of what lower-level encodings or physical devices are utilized, one may first proceed at the level of \textit{logical qudits}, where each $\integers_d$ is mapped to the basis states $\{\ket{0},\ket{1},\dots,\ket{d-1}\}$. This idea was applied, for instance, in various Quantum Alternating Operator Ansatz constructions of \cite{hadfield2019quantum}. 
QAOA for qudits has been further considered in \cite{bravyi2022hybrid,sawaya2022encoding,ozguler2022numerical,deller2023quantum,venturelli2025near,apte2026quantum}, among other works. 

As with qubits, different quantum circuit ans\"atze are possible for qudit-based approaches to combinatorial optimization. Here we briefly review QAOA, %
which consists of %
an initial state, along with %
phase and mixing operators which are applied in alternation, %
followed by a computational basis measurement to return a candidate problem solution. 
To implement QAOA at the logical qudit level we may consider, for instance, the initial state 
\begin{equation}
    \big( \frac1{\sqrt{d}}(\ket{0}+\ket{1}+\ket{2}+\dots+\ket{d-1})\big)^{\otimes n}
\end{equation}
which gives the usual equal superposition over all states; other initial states are possible~\cite{hadfield2019quantum}. 
The phase operator is 
$$U_P=exp(-i\theta H),$$
where $H$ is the cost Hamiltonian satisfying $H\ket{y}=c(y)\ket{y}$. 
For mixing operator $U_M$, different choices are possible~\cite{sawaya2022encoding}, in particular such that the %
design criteria of \cite{hadfield2019quantum} are satisfied. In general some operator of the form 
$$U_M=\bigotimes_{j=1}^n exp(-i \beta B_j)$$
suffices where each $B_j$ acts nontrivially on (typically) a single logical qudit. 
For unconstrained problems over integers, $B_j$ can be set to be (Hermitian combinations of) left- or right-shift operators, which generalize the usual transverse field $B_j=X$ mixer for qubits. 
For a $d$-level qudit %
we may generalize the qubit Pauli operators to become
the %
shift and clock operators
\begin{equation} \label{eq:quditPaulis}
X_d \ket{j} = \ket{j+1 \!\!\!\!\pmod d}, \quad
Z_d \ket{j} = \omega^j \ket{j}
\end{equation}
with $\omega_d = e^{2\pi i/d}$, 
which satisfy the Weyl commutation relation $ZX = \omega XZ$; when $d$ is clear from context we omit their subscripts. 
An important difference is that these operators are no longer Hermitian for $d>2$. Nevertheless a mixer can be constructed using $B_j=(X+X^\dagger)/2$ for each qudit~\cite{hadfield2019quantum,sawaya2022encoding}. 
We emphasize that the mixer $U_M$ may be specified and transformed at the logical level, and then mapped to appropriate gates implementing it once a hardware encoding is specified.

\section{NDAR for Qudits} \label{sec:ndarQudits}

Here we present our extension of NDAR to qudit-based quantum hardware, 
previously briefly teased in~\cite{venturelli2025near}. 
In what follows we significantly expand upon the prior work. 
We first motivate our construction by discussing how amplitude damping noise generalizes to qudit systems. 
For sufficiently low temperature the attractor state again approaches $\ket{00\dots 0}\bra{00\dots 0}$, while more generally it is a mixed thermal state dominated by %
low excitation number components.

\subsection{%
Generalized amplitude damping channels} \label{sec:genAmpDamp}
Consider for the moment a single system qudit Hamiltonian of the form
$\widetilde{H} = \sum_{j=0}^{d-1} E_j \ket{j}\bra{j}$, 
with $0=E_0 < E_1 < \cdots < E_{d-1}$; i.e., the spectrum of $\widetilde{H}$ defines the computational basis and is nondegenerate.  
For qubits ($d=2$), amplitude damping models energy relaxation
$\ket{1} \rightarrow \ket{0}$ 
with some probability~$p$ %
(e.g., spontaneous emission), which at zero temperature is described by the Kraus operators
\begin{equation}
    K_0=\ket{0}\bra{0}+\sqrt{1-p}\ket{1}\bra{1}, \quad
    K_1 = \sqrt p \ket{0}\bra{1}.
\end{equation}

For qudits, %
different generalizations are possible; for example, damping channels may allow limited or arbitrary downward jumps in energy. 
Generally amplitude damping means population flows downward in energy $\ket{d-1} \rightarrow \ket{d-2} \rightarrow \dots \rightarrow \ket{1}\rightarrow\ket{0}$. %
This type of channel is important for modeling various types of systems including for example leakage and relaxation in transmons, and loss in truncated bosonic channels. 
Generally, amplitude damping can result from coupling to a bath with modes that can absorb the qudit’s transition energy (electromagnetic modes, phonons, other two-level defects, etc.), turning excited-state energy into environmental excitations. This includes spontaneous emission into electromagnetic modes %
or phonons, %
dielectric loss from a bath of two-level systems, and thermal excitations, among other physical mechanisms.

\paragraph{Zero-temperature amplitude damping}
At zero temperature, the bath is its ground state (cannot provide energy), hence only downward jumps are allowed. %
Consider a single oscillator-like qudit and 
let $\eta=\textrm{exp}(-\kappa t)$ denote the transmissivity over the channel time $t$, so that $\gamma=1-\eta$ is the probability that each individual excitation is lost
(such as by single boson emission in harmonic oscillators). %
Then the number of lost excitations is binomially distributed, 
with the probability of the transition \(|j\rangle\to |j-i\rangle\)
given by %
$\binom{j}{i}(1-\gamma)^{j-i}\gamma^i$, 
and
so one defines the amplitude damping channel~\cite{chuang1997bosonic,grassl2018quantum,dutta2025noise} by the Kraus operators
\begin{equation} \label{eq:genAmpDamp}
    K_i = \sum_{j=i}^{d-1} \sqrt{\binom{j}{i}(1-\gamma)^{j-i}\gamma^i} \ket{j-i}\bra{j},
\end{equation}
for $i=0,1,\dots,d-1$, derived as a truncated bosonic pure-loss channel. 
In the short-time limit we have \(\gamma\simeq \kappa t\) and the dominant transition from %
each basis state is $|j\rangle\to |j-1\rangle$ with probability $j\kappa t+O(t^2)$, corresponding to adjacent transition rate $\Gamma_j^- = j\kappa$. 
Generally
this model allows all downward transitions,  
is non-unital, and has density matrix fixed
point $\ket{0}\bra{0}$. 
For $n$ qudits each under the action of this channel, the fixed point is then %
$\ket{00\dots 0}\bra{00\dots 0}$
in the sense that $\Lambda^\ell(\rho)\rightarrow \ket{00\dots 0}\bra{00\dots 0}$ 
as $\ell\rightarrow \infty$, for $\Lambda(\rho):= \sum_{i=0}^{d-1}K_i\rho K_i^{\dagger}$. 
Hence %
for qudits under %
the amplitude damping noise of Eq.~\eqref{eq:genAmpDamp}
we expect the output distribution to be skewed toward states of low excitation number, %
analogous to the skew towards states of low Hamming weight observed and exploited for qubits in~\cite{maciejewski2024improving}.

\paragraph{Finite-temperature generalizations}%
At finite temperature both absorption and emission are allowed, subject to detailed-balance conditions. This models a coupling to a thermal bath. 
From the continuous-time perspective, defining jump operators
$L_j^- = \ket{j-1}\bra{j}$ and $L_j^+ = \ket{j}\bra{j-1}$
the Lindbladian becomes
\begin{equation}\label{eq:Lindblad}
\mathcal{L}(\rho)
=
\sum_{j=1}^{d-1}
\left[
\Gamma_j^- \mathcal{D}[L_j^-](\rho)
+
\Gamma_j^+ \mathcal{D}[L_j^+](\rho)
\right],
\end{equation}
where
$ \mathcal{D}[L](\rho)
= L \rho L^\dagger - \frac{1}{2}\{L^\dagger L, \rho\}$. 
Assuming thermal equilibrium at inverse temperature $\beta$, from detailed balance the rates satisfy
$\Gamma_j^+/\Gamma_j^- = e^{-\beta (E_j - E_{j-1})}$ 
such that the unique fixed point is the Gibbs state
\[
\frac{e^{-\beta \widetilde{H}}}{\mathrm{Tr}(e^{-\beta \widetilde{H}})} 
= \frac1{\textrm{Z}}\bigg(\ket{0}\bra{0}+ e^{-\beta E_1}\ket{1}\bra{1}+ e^{-\beta E_2}\ket{2}\bra{2} +\dots \bigg).
\]
Hence, while the fixed point for each qudit is no longer $\ket{0}\bra{0}$, higher energy level components %
are exponentially suppressed with inverse temperature and energy. 

As a discrete channel over a time step $t$ the CPTP map is
$\Lambda_t = e^{t \mathcal{L}}$. Hence 
for small $t$, the Kraus operators are now approximately %
\[
\begin{aligned}
K_0 &\approx I -\frac{t}{2}\sum_{j=0}^{d-1} (\Gamma_j^- + \Gamma_{j+1}^+)\ket{j}\bra{j}, \\
K_j^- &\approx \sqrt{\Gamma_j^- t}\, \ket{j-1}\bra{j}, \quad 
K_j^+ \approx \sqrt{\Gamma_j^+ t}\, \ket{j}\bra{j-1},
\end{aligned}
\]
which generalizes the zero temperature case above in the short-time limit. 
(Here we take $\Gamma_0^-=\Gamma_{d}^+=0$, whereas more general %
device channels may also include leakage to non-computational states~\cite{wood2018quantification,li2025universal,kim2025ultracoherent}.) 
Hence, for $n$ qudits under finite-temperature amplitude damping noise we again expect the output distribution to be skewed toward low %
excitation states, and so in principle exploitable by NDAR. %

More generally still, one may consider distinct rates for each allowed transition between two states $\ket{i}$ and $\ket{j}$, such as for anharmonic spontaneous emission/absorption transitions in atomic systems~\cite{grassl2018quantum}.
In the short-time  
\sh{(i.e.,~$\Gamma_j^{\pm}t\ll1$)} 
limit 
single %
transitions between adjacent states in the ladder again dominate. %
In either setting the %
low-temperature bias toward low-energy labels remains. %

\subsection{NDAR for logical qudits}  \label{sec:ndarQuditsCriteria}

Suppose a problem on $\integers_d^n$ is mapped directly to $n$ $d$-level qudits 
(we sometimes call $\integers_d$ values \emph{colors}). 
We revisit the four ingredients of Sec.~\ref{sec:ndar} in this setting:
\begin{enumerate}
    \item \textbf{Attractor:} the qudit state $\ket{00\dots 0}$ remains the natural attractor, analogous to the qubit case. 
    
    This is motivated by the generalized amplitude damping channels of Sec.~\ref{sec:genAmpDamp}, which %
    act %
   to suppress excitation number on each qudit. Other attractors can be accommodated with minimal changes.
    
    \item \textbf{Feasibility:} the attractor encodes a feasible state for unconstrained problems, %
    as in the qubit case.

    \item \textbf{Gauge %
    selection:} the qubit one-to-one correspondence breaks, leaving an \emph{exponentially large family} of gauges compatible with each sampled string.
    
    For $d>2$, per-qudit relabelings of the $d$ logical levels yield $(d!)^n \gg d^n \gg 2^n$ gauge transformations. Even after fixing which level each qudit sends %
    $\ket{0}$ to, the remaining %
    $d-1$ levels can be permuted freely, leaving $((d-1)!)^n$ \emph{compatible} gauges per sampled string instead of a unique gauge as for qubits. 

    Given a target string $y\in\mathbb{Z}_d^n$ 
    we consider general gauge transformations
    \begin{eqnarray}
        H^{\y} 
    &:= P_{\y}^\dagger HP_{\y}
    \end{eqnarray}
    where $P_{\y}$ acts to map %
    the attractor state to~$\y$, defined here with the convention $P_{\y}\ket{0}=\ket{y}$ such that $\bra{0\dots 0} H^{\y} \ket{0\dots 0}=\bra{\y}H\ket{\y}$. We require  %
    that each $y$ determines $ P_{\y}$ efficiently and uniquely.

    \item \textbf{Practical circuit remappings:} the surplus gauge freedom becomes a design lever %
    toward hardware-friendly circuits: 

From the $((d-1)!)^n$ compatible gauge transforms one can \emph{choose} 
    $\{P_{\y}\}$ to be
    the cheapest or most convenient set to implement for the available hardware and ansatz (e.g.,\ transpositions swapping two levels, versus %
    cyclic shifts $X^r_d$ of all $d$ levels, for each qudit). %
    The goal is to use this freedom to select gauge transforms that preserve as much circuit structure as possible, e.g. by restricting to hardware-friendly subgroups (such as cyclic shifts). 
\end{enumerate}

Item 3 shows that qudits come with significantly more gauge freedom than qubits, which we argue is advantageous. In particular, this allows \emph{different sets of allowed gauge transformations to be tailored to different qudit encodings}. 
For example, for a single qudit, the previous and desired labels may be swapped, which doesn't affect the labels of any other levels, or a shift operator $X^r$ can be applied for $r$ the difference between the two values, which shifts the label of every qudit level. 
For hardware qudits, one can consider gauge transformations most compatible with the quantum ansatz employed as well as lower-level quantum gates. 
This freedom is especially important for the case of qudits encoded in qubits, where we show different encodings naturally lead to different choices of gauge transformations, distinct from each other as well as from the pure qudit encoding.

\begin{rem}
    The set of %
    gauge transforms often forms a \emph{subgroup} of the permutation group $(S_d)^n$ on encoded basis states, though not always. For qudit shift gauges of Eq.~\eqref{eq:quditshiftgauge} this subgroup is $\mathbb{Z}_d^n$ (acting
transitively on $\mathbb{Z}_d^n$ by translations). More general %
gauge transformation rules may select %
different representatives from the compatible elements of $(S_d)^n$ for each target string~$\y$.
\end{rem}

\subsection{Example: Graph $k$-coloring problems} \label{sec:quditkcol}
Here we consider graph coloring as a prototypical example of a problem over integers that captures the basic essence of scheduling, planning, and asset allocation problems. %
Deciding whether a given graph is properly $k$-colorable for $k\geq 3$ is a well-known NP-complete decision problem, and a variety of distinct optimization variants of this problem exist; see e.g. \cite{hadfield2019quantum,ausiello2012complexity}.

Given an $n$-node graph $G=(V,E)$ and $k$ colors $1,\dots, k$ (sometimes %
labeled $0,\dots, k-1$), find an assignment of colors to each node such that a particular cost function is optimized or satisfied, along with possible hard constraints that must be satisfied.   
Different problem variants exist related to both proper and improper colorings. A coloring is proper if %
no edge in the graph connects vertices of the same color, else improper.  %
Here we use coloring to denote any possible color assignment.

\paragraph*{Max-$k$-colorable subgraph problem \cite{hadfield2019quantum,bravyi2022hybrid}} given fixed $k\geq 2$, maximize the objective function
$$c(x) = \# \textrm{ edges connecting differently colored vertices}$$
for all possible assignments $x \in \{1,2,\dots,k\}^n$ of $k$ colors to $n$ vertices. 
This problem naturally generalizes Maximum Cut, %
a commonly considered application problem for quantum algorithms, %
\sh{reducing to it exactly when $k=2$.}
Applications related to scheduling problems include variants where each node may have its own set of allowed colors, and the goal remains to minimize conflicts. Here for simplicity we consider the version as described.

\paragraph*{Qudit encoding}
Suppose we have available a quantum computer with $n$ logical $k$-level qudits (i.e., $d=k$).
First rewrite the cost function %
as
\begin{eqnarray*}
    c(x) &=&\sum_{(uv)\in E} \text{NEQ}(col_u(x),col_v(x)) .
\end{eqnarray*}
Here $col_u(x)$ denotes the function that returns the color assigned to a vertex $u$ by a string %
$x\in \mathbb{Z}_k^n$, %
and NEQ denotes the %
Not Equal function. %
We identify these strings with basis states, %
i.e.,  $\ket{red},\;\ket{green},\;\ket{blue},\dots,\ket{violet}$, 
which we conveniently identify with integers and the qudit states 
\begin{equation} \label{eq:basis_states}
    \ket{0},\ket{1},\ket{2},\dots,\ket{d-1}.
\end{equation}
In this representation the cost Hamiltonian becomes 
\begin{align*}
    H = |E|I-\sum_{(uv) \in E}&\big (\ket{00}\bra{00}+\ket{11}\bra{11}+\ket{22}\bra{22}+\dots\\ &+ \ket{(d-1)(d-1)}\bra{(d-1)(d-1)}\big)
\end{align*}\label{eq:projection_primitives}
Here we represent the cost Hamiltonian using projector primitives~\cite{sawaya2022encoding} $\Pi_{u,v} := \sum_{j=0}^{d-1} \ketbra{j}{j}_u \otimes \ketbra{j}{j}_v = \sum_{j=0}^{d-1} \ket{jj}_{uv} \bra{jj}_{uv}$. %
For qudits, a variety of operator bases to represent such operators and ultimately quantum gates are possible~\cite{wang2020qudits,sawaya2022encoding}.

\paragraph*{Qudit-based ansatz}
Given $H$, we can construct an ansatz such as QAOA %
(Sec.~\ref{sec:QAOAqudits}),  
which may be compiled to 
logical or low-level
qudit gates in different ways. %

\paragraph{Compilation via generalized Pauli operators}
Recall Eq.~\eqref{eq:quditPaulis} above. 
On $\mathbb Z_d$ we have the identity
$\delta_{x_u,x_v} =\frac{1}{d}\sum_{m=0}^{d-1} \omega^{m(x_u-x_v)}$, 
and so observing %
$Z_u^m Z_v^{-m}\ket{j_u,j_v}=\omega^{m(j_u-j_v)}\ket{j_u,j_v}$
we have
$\sum_{j=0}^{d-1}\ket{jj}\bra{jj}_{uv}=
\frac{1}{d}\sum_{m=0}^{d-1} Z_u^m Z_v^{-m}$. 
Hence the cost Hamiltonian %
becomes 
\begin{equation} \label{eq:costHamqudits}
H = (1-\frac1d)|E|I-\frac{1}{d}\sum_{(u,v)\in E}\sum_{m=1}^{d-1} Z_u^m Z_v^{-m},
\end{equation}
where a Hermitian decomposition is obtained by observing that all terms occur in conjugate pairs $Z_u^m Z_v^{-m}+Z_u^{-m} Z_v^{m}$, 
and so the phase separator
$e^{-i\gamma H}$ decomposes into commuting edge phases implementable with
two-qudit controlled-phase gates.%

For the mixer, %
the transverse field Hamiltonian %
generalizes to a Hermitian combination of shift operators
\begin{equation} \label{eq:quditMixer}
B = \sum_{v\in V} (X_v + X_v^\dagger)
\end{equation}
of Eq.~\eqref{eq:quditPaulis}. 
As $[X,X^\dagger]=0$, the mixer $e^{-i\beta B}$ can be implemented exactly as the product of partial mixing terms $e^{-i\beta (X+X^\dagger)}$. 
The partial mixers %
are generally not Clifford operations, except at special angles or for special $d$.  
Nevertheless they can be easily compiled from the Fourier transform identity
$F X F^\dagger = Z$
which implies 
$X+X^\dagger$ is diagonalized by  $F^\dagger$.
In particular, if one can implement a diagonal unitary in the $Z$ basis %
then a partial mixing term can be implemented by conjugation as
\begin{equation*}
e^{-i\beta (X+X^\dagger)}
=
F^\dagger \left(e^{-i\beta (Z+Z^\dagger)}\right) F.
\end{equation*}

\paragraph{Compilation to SUM and controlled-phase}
Consider the two-qudit controlled-phase gate which acts as 
$\mathrm{CP}\ket{a,b} = \omega^{ab}\ket{a,b}$.
This is a Clifford gate for prime~$d$, and is related to 
the qudit SUM gate, SUM$\ket{a,b}=\ket{a,b+a}$, by conjugation with $F$ on one qudit, $CP=(I\otimes F)\textrm{SUM}(I\otimes F^\dagger)$. 
Using these primitives
and single-qudit gates%
~\cite{%
proctor2017ancilla,murairi2024highly,yang2025quantum} 
one can implement $e^{-i\theta (Z_u^m Z_v^{-m} +Z_u^{-m} Z_v^{m})}$ with %
standard phase kickback: 
compute the modular difference into an  ancilla register using SUM and inverse-SUM, apply a single-qudit diagonal phase gate on that register, and  uncompute. %

\paragraph{Compilation to hardware} 
Different low-level gate sets and primitives exist to %
compile the above logical operators to. SNAP and Displacement gates~\cite{krastanov2015universal,heeres2015cavity} are universal for single qudit operations, and can implement multiqubit operations once augmented with entangling gates~\cite{job2023efficient, kurkccuoglu2024qudit, ogunkoya2025investigating}. 
Other %
options include echoed conditional displacement gates~\cite{eickbusch2022fast,diringer2024conditional,lapointemajor2024optimizing} which use an ancilla qubit dispersively coupled to an oscillator or cavity, as well as 
optimized 
direct control-level pulses~\cite{ozguler2022numerical,ozguler2024dynamics}. 
We provide some further %
discussion in Sec.~\ref{sec:discussion}.

\paragraph*{Gauge transform fixing} Starting at the logical qudit level we propose the choice of gauge transformation  
\begin{eqnarray} \label{eq:quditshiftgauge}
    V(r) = X^{r_1}X^{r_2}\dots X^{r_n}
\end{eqnarray}
where $X:=X_d$ is the left-shift operator and 
$r_u=col_u(x^*)-col_u(0)$ 
is defined for each qudit (integer variable) as the difference between the integer label of the best solution found at the current step and that encoded by~$0$. %
Under this choice %
the cost Hamiltonian \eqref{eq:costHamqudits} transforms under $V=V(r)$ as 
\begin{equation*}
V^\dagger H V
=(1-\tfrac{1}{d})|E|I
-
\frac{1}{d}
\sum_{(uv)\in E}\sum_{m=1}^{d-1}
\omega^{m(r_u-r_v)}
Z_u^m Z_v^{-m}.
\end{equation*}
Hence we see that here each Pauli term in $H$ may be phase shifted by a $d$th root of unity, generalizing the sign flip case for qubits. (Similar formulas apply for higher order Pauli $Z_d$ terms.)
This choice is natural for many qudit gate sets. As we will see below, different sets of gauge transformations %
will be convenient when the same problem is mapped to qubits. %

\section{NDAR for encoded variables} \label{sec:ndarencoded}

Here we extend NDAR-QAOA to integer domains encoded with qubits, and consider tradeoffs in terms of suitability for NDAR %
as compared to the qudit case presented above. To illuminate these results we continue with the Max-k-colorable subgraph problem as our running example. %
See e.g. \cite{hadfield2019quantum,chancellor2019domain} for discussion of encodings and tradeoffs. 
Here we focus on the main differences from the results given in Sec.~\ref{sec:ndarQudits}.

\subsection{Binary qubit encoding}

In binary encoding we use $\ell =\lceil \log_2(k) \rceil$ qubits per vertex to encode each color. 
Here we identify the logical basis states of Eq.~\eqref{eq:basis_states} with the 
$\ell$-qubit basis states %
\begin{equation}  \label{eq:basis_states_bin}
    \ket{0\dots 00}, \ket{0\dots 01} , \ket{0\dots 10} , %
    \dots , \ket{1\dots 11},
\end{equation}
with %
$\mathcal{E}(j)=|j\rangle_2 = |j_{\ell-1}\cdots j_1 j_0\rangle$, $j=\sum_{i=0}^{\ell-1} j_i 2^i$. 
Here we assume $k$ is a power of $2$, %
\sh{so all possible states are valid,} and in particular the standard transverse-field mixer 
\begin{equation} 
U_M = \prod_{v=1}^n \prod_{j=0}^{\ell-1} \exp(-i \beta X_{v_j})
\end{equation}
is appropriate %
and can be implemented with $\ell n$ single qubit ($X$-rotation) gates.
(If $k\neq2^\ell$, one could assign multiple strings per color, resulting in more complicated cost and mixing operators, or apply other strategies with different tradeoffs~\cite{hadfield2019quantum,fuchs2021efficient,sawaya2022encoding,fuchs2025encodings}.) %

For the initial state,
$ \ket{\psi_0}=\ket{+}^{\otimes \ell n}$
encodes the equal superposition over all possible coloring assignments and is trivially prepared.
The cost Hamiltonian %
becomes
\begin{equation} \label{eq:costHamBin}
    H=|E|I - \sum_{(uv)\in E} \sum_{j=0}^{k-1} \ket{\mathcal{E}(j)\mathcal{E}(j)}\bra{\mathcal{E}(j)\mathcal{E}(j)}
\end{equation}
which can be uniquely %
mapped to Pauli $Z$ qubit operators~\cite{hadfield2021representation}. 
For example, for $k=4$ this becomes 
    $\sum_{(uv) \in E} (\frac34 I - \frac14 Z_{u_0}Z_{v_0} - \frac14 Z_{u_1}Z_{v_1} - \frac14 Z_{u_0}Z_{v_0}Z_{u_1}Z_{v_1})$.
For $k=2^\ell$ %
 this generalizes to $k$ terms per edge and $H$ becomes $2\ell$-local (i.e., its terms act on up to $2\ell$ qubits). 

For the phase operator $U_P=exp(-i\gamma H)$, dropping the constant term and utilizing commutativity gives a product of $k-1$ multiqubit Pauli rotations. In the standard decomposition of  \cite{barenco1995elementary} %
this requires at most $|E|(k-1)$ single qubit rotations and $O(|E|\ell k)$ CNOT gates; other gate sets and optimizations are possible. 
Hence we see the resource requirements for QAOA for $k$-coloring in binary encoding are modest; in particular %
for $k=4$ they are roughly double that of QAOA for MaxCut.

\subsubsection*{NDAR for binary encodings} 
Suppose we have used QAOA %
or some other suitable algorithm for obtaining a set of %
solution samples, and assume that %
the system noise is 
biased towards an attractor state. 
For the binary encoding we assume the global attractor to be $\ket{00\dots 0}_{\ell n}$.

In this encoding, permutation of the integer label of a given variable corresponds to permutation of the %
computational basis vectors such %
that the set of gauge transformations is given by an $n$-fold tensor product of %
$2^\ell\times 2^\ell$ permutation matrices. 
Arbitrary basis permutations can be compiled into reversible classical circuits, generally requiring multi-controlled gates or their decompositions. 
However, the subgroup of \sh{invertible} affine transformations ($x\mapsto Ax+b$) can always be decomposed into %
$X$
and CNOT gates (after noting the necessary SWAP operation decomposes into 3 CNOTs), which can help satisfy the requirement of circuit remappings remaining practical.

Indeed, arbitrary %
permutations as gauge transformations may change the form of the resulting cost Hamiltonian, and the resulting circuit for %
algorithms such as QAOA. For example, consider the gauge transformation~$U=\text{SWAP}_{12}$ applied to a binary register. A term $Z_2Z_3$ in the cost Hamiltonian will transform as $U^\dagger Z_2Z_3 U = Z_1Z_3$, which is more than just a coefficient sign change, %
and may for example introduce %
swaps or other changes to the hardware-level quantum circuit.

Fortunately, 
in binary encodings there always exists a set of gauge transformations of size~$k^n$ 
(i.e., one-to-one with integer strings)
that only affects the signs of terms in the cost Hamiltonian -- these are precisely those generated by products of $\ell n$ %
bitflip Pauli~$X$ operators that correspond to flipping the bits that differ between the physical attractor and encoded target string.  
Logically, this amounts to the affine transformation $x\mapsto x\oplus y$ bitwise, %
represented by the qubit gauge transformation
\begin{eqnarray} \label{eq:gaugeXfmBin}
    V(y)= \bigotimes_{v=1}^n \bigotimes_{j=0}^{\ell-1}X^{y_{v_j}}_{v_j},
\end{eqnarray}
where the variable %
$y_{v_j}$ is the difference between the $j$th color bit for vertex $v$ in $y$ and that of the attractor. 
This shows that we can always pick a compatible gauge transformation that decomposes into purely bit flips. 
The same logic as for bitflip transforms in the binary case %
then %
implies that this choice of $V(y)$ leaves the cost Hamiltonian the same up to signs of terms as desired.

\emph{Outlook:}
Integer problems encoded in binary appear compatible and attractive for NDAR as our analysis demonstrates, %
especially %
when $k$ is a power of 2. %
\sh{%
Our construction also applies 
to full Hilbert-space encodings for the $k\neq 2^\ell$ case 
in which every computational basis state is assigned to a valid color~\cite{fuchs2021efficient,fuchs2025encodings}.} 
Binary encodings generally %
facilitate space-efficient representation of the problem domain.
The primary tradeoff is the requirement of $2\lceil \log_2 k \rceil$-local terms in the cost Hamiltonian and phase operator; fortunately, compilations to $2$-local gates exist~\cite{barenco1995elementary}, and indeed this requirement has not been a major limitation in other results for 
quantum gate-model approaches~\cite{hadfield2019quantum,shaydulin2024evidence}.

\begin{rem}
    Equation~\eqref{eq:basis_states_bin} considers a standard binary encoding; the %
    insights of this section similarly apply to other related variants such as Gray codes~\cite{di2021improving,sawaya2022encoding}.
\end{rem}

\subsection{Unary qubit encodings}

Here we %
contrast %
three variants of unary encodings, which require $\sim k$ qubits per integer variable. These encodings treat different $k$ values uniformly, but come with additional (exponentially larger) %
qubit overhead.

\subsubsection{One-hot encoding}
In standard one-hot encoding we use $k$ qubits per vertex to encode each color. 
We identify the $k$ logical basis states of Eq.~\eqref{eq:basis_states} with the $k$ basis states of Hamming weight one 
\begin{equation}  \label{eq:basis_states_onehot}
    \ket{10\dots 00}, \;\ket{01\dots 00} , \;\dots ,\;\ket{00\dots 01}.
\end{equation}

Restricted to this valid %
subspace %
the cost Hamiltonian is 2-local and can be expressed %
\cite[Eqn. 6.22]{hadfield2018quantum} as
\begin{equation}
    C =\; \frac{k|E|}4I-\frac14 \sum_{(uv) \in E} \sum_{j=0}^{k-1} Z_{u,j}Z_{v,j}.
\end{equation}

QAOA for this problem and encoding is %
detailed in~\cite{hadfield2019quantum}. The phase operator decomposes into at most $2k|E|$ CNOT gates. For this encoding, qubit XY mixers of the form %
$U_M(\beta)=\prod_{a=1}^{n} U_{M,a}(\beta)$ with 
$$U_{M,a}(\beta)=\prod_{(ij)\in S(a)} exp(-i\beta X_iX_j)exp(-i\beta Y_iY_j)$$
are proposed, which preserve qubit Hamming weight for each of the encoded qudits and hence never map valid to invalid states. Here the possible index set $S(a)$ for the qubits of each encoded variable includes both ring and fully-connected topologies, among others~\cite{hadfield2019quantum, wang2020xy}.

\paragraph*{NDAR} A practical difficulty with this scheme is that noise in general will cause leakage of probability amplitude out of the feasible subspace, and hence some fraction of measurement outcomes will yield invalid strings.
For each logical qudit, the valid states are an exponentially small fraction of the encoded Hilbert space with respect to increasing $k$. Hence an attractor state is unlikely to correspond to a valid solution. 
(If the attractor was known to be a %
basis state of Eq.~\eqref{eq:basis_states_onehot}, then this case is logically equivalent to the qudit encoding of Sec.~\ref{sec:ndarQudits}, and our criteria would be satisfied so that we could proceed with NDAR as usual.)
Indeed, our motivating attractor $\ket{000...000}$ is invalid for standard one-hot encodings. 
The next encoding we consider augments %
the logical one-hot subspace to include the $\ket{00\dots0}$ state.

\subsubsection{Augmented ``01-hot" encoding}
Consider a variant of one-hot encoding where we use $k-1$ qubits per vertex instead of $k$ to encode each color by also including the $\ket{00\dots 0}$ state to encode the first color. Such an encoding was considered in \cite{hadfield2019quantum}. 
Here we identify the logical basis states of Eq.~\eqref{eq:basis_states} with the $1+(k-1)$ qubit basis states 
\begin{equation*}  \label{eq:basis_states_onehot2}
    \ket{00\dots 00}, \;\ket{10\dots 00}, \;\ket{01\dots 00} , \;\dots, \;\ket{00\dots 01}.
\end{equation*}

\paragraph*{NDAR} 
The color corresponding to $\ket{00\dots0}$ is indicated by the projector 
$2^{-(k-1)}\prod_{i=0}^{k-2}(I+Z_{v_i})$,
which contains a number of terms that grows exponentially with $k$ and has locality up to $k-1$. 
Hence detecting (in)equality of two zero-color states introduces the product of indicators
over both registers, yielding up to $2(k-1)$-local terms for the cost operators in general. 
On the other hand, one can always compile phase operators by again computing string inequality in an ancilla scratchpad register, performing phase kickback, and then uncomputing the ancilla qubits~\cite{barenco1995elementary,hadfield2021representation}.

Hence, in this encoding, the infeasible attractor issue goes away, and one can proceed with NDAR at the meta-algorithm level. 
The catch is that the cost Hamiltonian and corresponding phase operator transform in a nontrivial way and are no longer $2$-local.
Hence this encoding may be suitable for, %
e.g., hardware-efficient ans\"atze (that don't depend on $H$),
but less so for QAOA.

\subsubsection{Domain-wall encoding}
Consider instead the \textit{domain-wall} encoding of \cite{chancellor2019domain} where we also use $k-1$ qubits per vertex to encode each color. 
Here we identify the logical basis states of Eq.~\eqref{eq:basis_states} with the $k$ basis states 
\begin{equation*}  \label{eq:basis_states_dw}
    \ket{000\dots 00}, \;\ket{100\dots 00}, \;\ket{110\dots 00} , 
     \dots ,
    \;\ket{111\dots 11},
\end{equation*}
i.e., $k$ colors are encoded in $k-1$ qubits as $r\rightarrow \mathcal{E}(r):=\ket{1^r0^{k-1-r}}$, $r=0,1,\dots,k-1$. 
Here, 
again assuming we are restricted to the subspace of valid states, 
value indicator operators correspond to terms of the form $Z_i-Z_j$ or $I-Z_i$, such that %
$2$-local qudit cost Hamiltonians are mapped to 2-local (QUBO) Hamiltonians %
and phase operators on qubits. 
Explicit derivations are shown in \cite{chancellor2019domain} by considering the action on the logical subspace.  
For QAOA mixers, Hamiltonian terms of the form $Z_iX_j-X_iZ_j$ and $\pm X_j$ are shown to be sufficient (in the sense of the QAOA mixer design criteria of \cite{hadfield2019quantum}). %

\paragraph*{Gauge transforms}
Here we propose the prefix-flip gauge as a natural choice: for target $\y\in \mathbb{Z}_k^n$ let 
\begin{eqnarray}
    P_{\y}:=\bigotimes_{i=1}^n \bigotimes_{j=1}^{y_i} X_{i,j}.
\end{eqnarray}
It is easy to verify this satisfies $P_{\y}\ket{00\dots 0}=\ket{\mathcal{E}(\y)}$ as desired. 
On Pauli strings, prefix-flip gauges again act only by sign changes.  %
Hence a domain-wall cost Hamiltonian
retains the same Pauli support and
locality after remapping. However, individual domain-wall value indicators,
being linear combinations of $I$ and $Z_i$, are not generally multiplied
by an overall sign nor mapped to identical qubit operators %
but may transform to other diagonal operators of the same locality.

\paragraph*{NDAR} In this encoding the %
presumed physical qubit attractor state $\ket{00\dots 0}$ is always valid, 
and one can proceed with NDAR at the meta-algorithm level.

\emph{Outlook:} 
Of the unary encodings considered, domain-wall %
presents the best suitability for NDAR %
with advantages over both one-hot encoding and the 01-hot variant. 
Nevertheless for all unary encodings noise is likely to lead to invalid strings, and hence they remain less attractive than qudit-native or binary encodings. %

\subsection{Tradeoffs between quantum encodings}

We summarize the tradeoffs between the different encodings considered 
for Max-$k$-colorable subgraph 
in Table~\ref{tab:summary} above.  
The top of the table contrasts qudit encodings with binary and unary qubit encodings in terms of resources and general suitability for NDAR. While both one-hot and domain-wall encodings always produce feasible solutions in the ideal limit (assuming a feasibility-preserving quantum circuit~\cite{hadfield2019quantum}), in practice noise and other types of errors will reduce this probability to be less than one. Binary encodings, on the other hand, have the desirable property that all strings are feasible, natively when $k$ is a power of $2$, \sh{and through other %
modifications more generally~\cite{hadfield2019quantum,fuchs2021efficient,fuchs2025encodings}.} %
For incorporating NDAR, four out of the five encodings considered have the desirable property that the $\ket{00\dots 0}$ state corresponds to a valid assignment. %

The bottom two rows of the table %
show tradeoffs for applying an algorithm like QAOA using these encodings. While a binary encoding uses less space, it requires higher locality operators for the cost Hamiltonian and phase operator than the unary encodings, \sh{with} 
additional complications %
when $k$ is not a power of $2$. 
For arbitrary $k$ the locality becomes $2\lceil\log_2 k\rceil$, %
whereas the unary encoding cost operators remain 2-local. 
Note that on gate-model devices arbitrary $\ell$-local interactions can be compiled to two-qubit gates, sometimes by incorporating additional ancilla qubits, but with circuit depth proportional to $\ell$~\cite{barenco1995elementary}. 
Still, the case is not so clear-cut as all encodings have a number of terms that scale linearly with~$k$.

The observations here illustrate that tradeoffs between different encodings are not always so clear-cut, in particular for problems with modest $k$ or problem size, and especially when one seeks to incorporate NDAR. This is reflected in the literature where different encodings are employed in different use cases. Hence, our results above may prove useful for incorporating NDAR across a wider variety of experiments in the near term. %

\section{Discussion}
\label{sec:discussion}

In this work we extended the NDAR meta-algorithm beyond unconstrained binary
optimization %
to integer-domain %
problems on both qudit-native
and qubit-encoded hardware, including classes of problems with hard constraints. Treating both
axes within a single framework opens NDAR to substantially broader applicability
--- and, critically, to a substantial fraction of emerging quantum hardware
beyond the qubit paradigm. Our prototypical example, graph %
$k$-coloring,
already illustrates the central tension between encodings: qudit-native ($d=k$) and
qubit-encoded realizations are theoretically close cousins
but differ qualitatively in feasibility, gauge
structure, and circuit cost when integrated with NDAR. 
As summarized in Table~\ref{tab:summary}, we find
qudit-native encodings to be the %
best fit when available. %
For qubits, while binary encodings appear favorable in particular cases,  domain-wall encodings are applicable for all $d$ and present relative advantages over the other %
qubit encodings considered in terms of NDAR.

Three findings drive most of our conclusions. First, NDAR's four ingredients
(Sec.~\ref{sec:ndar}) are %
aligned natively, not incidentally, %
with 3D
superconducting cavity-based qudit processors: cavity photon loss is
well-modeled by 
the generalized amplitude damping channel, %
so the noise picture underlying NDAR appears to be  
well-matched to actual device noise~\cite{kim2025ultracoherent},
not 
purely an idealization. Second, the 
$d$-level 
qudit gauge
multiplicity $((d-1)!)^n$ is a design lever rather than a defect: among
compatible gauge transformations, one can \emph{select} the cheapest to
implement on the available hardware~\cite{kurkccuoglu2024qudit}, a freedom absent in the qubit case. 
Third, among the encodings we considered,
only the qudit-native encoding 
gives an \emph{always-feasible}
attractor 
\sh{independently of~$d$,}
with no encoding-induced invalid subspace \sh{or need for additional processing}, 
placing it on a
different footing from binary, one-hot, and domain-wall qubit encodings even
when their bare circuit costs may otherwise look comparable.

This paper is the algorithmic follow-up to a roadmap recently put forward
in~\cite{venturelli2025near}, where NDAR-QAOA on graph coloring was identified
as the near-term optimization workload best matched to emerging
superconducting cavity qudit processors~\cite{su2021construction,
Blok2021QutritScrambling, chi2022programmable, Ringbauer2022UniversalQudit,
roy2023two, denys2023, nguyen2024empowering, Gokhale2019QutritCircuits}. Two
open theoretical questions were flagged there: how NDAR generalizes when the
bitflip group is replaced by a richer gauge structure, and how qudit-native
encodings compare to qubit encodings of the same integer problem. The
framework of Secs.~\ref{sec:ndarQudits}--\ref{sec:ndarencoded} answers both.
The hardware side has moved in parallel. The SQMS Center's recently reported
two-mode TESLA-geometry processor~\cite{reineri2023exploration} 
demonstrates the
NDAR-relevant regime concretely, with single-photon lifetimes
$T_1 \simeq 20\,\mathrm{ms}$, Ramsey $T_2 > 20\,\mathrm{ms}$, and Fock-state
preparation up to $\ket{20}$ at fidelity above $95\%$~\cite{kim2025ultracoherent} via sideband feedforward
and parity-filter post-selection. %
Algorithm-agnostic
benchmarks~\cite{bornman2025benchmarking} corroborate that contemporary
transmons can drive a few tens of Fock levels at these coherence parameters,
placing a useful working bound $d \lesssim 20$ on the per-mode dimension
addressable today. A %
hardware demonstration of NDAR-QAOA
nevertheless requires two R\&D 
pushes: %

\emph{(i) High-fidelity two-qudit entangling gates} in the cavity-transmon
architecture~\cite{liu2024hybrid, ozguler2022numerical}. %
The
controlled-SUM (CSUM) gate is the natural phase-separator primitive but
remains the central synthesis-and-calibration challenge on multimode
cavities. Several primitive families are converging on this target. On the
Gaussian side, parametric beamsplitter gates between cavity modes have
reached $99.92\%$ raw ($99.98\%$ leakage-detected) fidelity using
parity-protected couplers~\cite{lu2023high}, and ancilla-mediated extensions
promote them to entangling gates with simulated error rates below
$10^{-4}$~\cite{tsunoda2023error}. Sideband-based control of multimode
memories has been pushed $30\times$ beyond the dispersive limit,
enabling fast cavity--cavity SWAPs in $10$-mode devices with millisecond-scale
photon lifetimes~\cite{huang2025fast}. Most directly relevant to NDAR-QAOA,
the same SQMS platform of~\cite{kim2025ultracoherent} provides an explicit
qutrit CSUM blueprint synthesized from five virtual-Raman beamsplitter
operations together with single-qutrit rotations, with simulated process
fidelity $\sim\!99\%$. A complementary universal primitive in the
weak-dispersive regime is Echoed Conditional Displacement
(ECD)~\cite{eickbusch2022fast}, whose multimode generalization, the
Conditional-NOT-Displacement (CNOD) gate, entangles modes of a single
multimode cavity through one weakly coupled transmon roughly two orders of
magnitude faster than the dispersive limit~\cite{diringer2024conditional}, with
optimal-control pulse shaping further reducing residual coherent
errors~\cite{lapointemajor2024optimizing}. ECD-based variational ans\"atze
have already been demonstrated experimentally on cavity-transmon hardware~\cite{dutta2025simulating}, providing
a concrete template for how %
NDAR-QAOA 
could be compiled in
practice. Crucially, the qudit mixer of Eq.~\eqref{eq:quditMixer} often maps onto
\emph{unconditional} cavity displacements that bypass the transmon ancilla
altogether, while the diagonal phase separator is generated by the natural
cross-Kerr evolution at (in principle) no gate cost~\cite{ozguler2022numerical}. Combined
with CSUM-style entanglers for non-quadratic (clock-shift) cost terms,
ECD/CNOD and beamsplitter primitives thus form a near-complete native gate
set for QAOA on Fock-encoded cavity qudits.

\emph{(ii) Error suppression and gauge-aware compilation.} NDAR is beneficial
when the dominant noise has a stable, characterizable bias. Dephasing-dominated or higher-temperature regimes weaken the attractor picture, and on
encoded variables noise can cause leakage from the feasible subspace. %
Suppressing non-damping
components of device noise via dynamical-decoupling sequences, randomized
compiling, or partial error-correction on either the transmon or cavity modes~\cite{tripathi2025qudit, suri2024uniformly,
dambal2025harnessing} is %
hence complementary %
to NDAR, 
particularly given recent evidence that NDAR can be negatively impacted by
residual dephasing~\cite{tam2025enhancing, newpaperunpublished}. Quantum
error correction 
and fault-tolerant 
\sh{technologies} 
for qudits remain active adjacent
areas offering further 
\sh{%
refinements and}  
sophistications~\cite{Knill1996NonBinary, Gottesman1999FaultTolerant,
Ashikhmin2001Nonbinary, Hostens2005Stabilizer,bullock2007qudit,godwood2026fault}.

Beyond these two %
priorities the framework opens broader directions.
While here we focused on integer domains and constraints induced by encoding, NDAR appears %
amenable to
extensions to other discrete domains (e.g.\ permutations, for problems such as
the Traveling Salesperson) and %
further tailoring to specific
classes of hard constraints. 
In closely related recent work~\cite{newpaperunpublished} we have developed
a new meta-algorithm paradigm that combines NDAR with iterative warm-starting~\cite{%
egger2021warm, 
lopez2025non, marshall2026quantum,
lotshaw2026iterative}; the results of
this paper port directly to that setting. 
Other recent follow-on work has shown
that further improvements to NDAR are possible via incorporating 
additional classical
processing~\cite{maciejewski2024multilevel, tam2025enhancing}, and these
insights should similarly carry over. More speculatively, %
better leveraging open-system and thermodynamic considerations, whether 
algorithmic~\cite{diez2023quantum, lotshaw2023approximate,stollenwerk2025measurement} or physical
(Sec.~\ref{sec:genAmpDamp}), is an attractive direction for 
further  
principled
algorithm modifications. 

\section*{Acknowledgments} 
 We %
 are grateful for 
helpful discussions with Tanay~Roy, Silvia Zorzetti, Doga Kürkçüoğlu, %
and %
Daniel Koch.
We acknowledge support from NASA (ISRDS contract 80ARC020D0010) and Department of Energy, Office of Science, National Quantum Information Science Research Centers, Superconducting Quantum Materials and Systems Center (SQMS), under Contract No. 89243024CSC000002, and grant No. DE-SC0026126; and from AFRL Contract No. FA8750-25-C-B0040.

\bibliographystyle{ieeetr}
\bibliography{bib}

@article{hadfield2019quantum,
  title={From the quantum approximate optimization algorithm to a quantum alternating operator ansatz},
  author={Hadfield, Stuart and Wang, Zhihui and O’Gorman, Bryan and Rieffel, Eleanor G and Venturelli, Davide and Biswas, Rupak},
  journal={Algorithms},
  volume={12},
  number={2},
  pages={34},
  year={2019},
  publisher={MDPI}
}

@article{hadfield2021representation,
  title = {On the Representation of {B}oolean and Real Functions as {H}amiltonians for Quantum Computing},
  volume = {2},
  ISSN = {2643-6817},
  url = {http://dx.doi.org/10.1145/3478519},
  DOI = {10.1145/3478519},
  number = {4},
  journal = {ACM Transactions on Quantum Computing},
  publisher = {Association for Computing Machinery (ACM)},
  author = {Hadfield,  Stuart},
  year = {2021},
  month = dec,
  pages = {1–21}
}

@article{sawaya2022encoding,
  doi = {10.22331/q-2023-09-14-1111},
  url = {https://doi.org/10.22331/q-2023-09-14-1111},
  title = {Encoding trade-offs and design toolkits in quantum algorithms for discrete optimization: coloring, routing, scheduling, and other problems},
  author = {Sawaya, Nicolas PD and Schmitz, Albert T and Hadfield, Stuart},
  journal = {{Quantum}},
  issn = {2521-327X},
  publisher = {{Verein zur F{\"{o}}rderung des Open Access Publizierens in den Quantenwissenschaften}},
  volume = {7},
  pages = {1111},
  month = sep,
  year = {2023}
}

@article{egger2021warm,
  author = {Egger, Daniel J and Mare{\v c}ek, Jakub and Woerner, Stefan},
  doi = {10.22331/q-2021-06-17-479},
  journal = {Quantum},
  pages = {479},
  publisher = {Verein zur F{\"o}rderung des Open Access Publizierens in den Quantenwissenschaften},
  title = {{W}arm-starting quantum optimization},
  volume = {5},
  year = {2021}
}

@article{lopez2025non,
  title={A Non-Variational Quantum Approach to the Job Shop Scheduling Problem},
  author={Lopez-Ruiz, Miguel Angel and Tucker, Emily L and Arnold, Emma M and Epifanovsky, Evgeny and Kaushik, Ananth and Roetteler, Martin},
  journal={arXiv preprint arXiv:2510.26859},
  year={2025}
}

@article{maciejewski2024improving,
  title={Improving quantum approximate optimization by noise-directed adaptive remapping},
  author={Maciejewski, Filip B and Biamonte, Jacob and Hadfield, Stuart and Venturelli, Davide},
  journal={Quantum},
  volume={9},
  pages={1906},
  year={2025},
  publisher={Verein zur F{\"o}rderung des Open Access Publizierens in den Quantenwissenschaften}
}

@article{tam2025enhancing,
  title={Enhancing {NDAR} with Delay-Gate-Induced Amplitude Damping},
  author={Tam, Wai-Hong and Matsuyama, Hiromichi and Sakai, Ryo and Yamashiro, Yu},
  journal={arXiv preprint arXiv:2504.12628},
  year={2025}
}

@article{venturelli2025near,
  title={Near-term Application Engineering Challenges in Emerging Superconducting Qudit Processors},
  author={Venturelli, Davide and Gustafson, Erik and Kurkcuoglu, Doga and Zorzetti, Silvia},
  journal={arXiv preprint arXiv:2506.05608},
  year={2025}
}

@article{lucas2014ising,
  title={Ising formulations of many {NP} problems},
  author={Lucas, Andrew},
  journal={Frontiers in Physics},
  volume={2},
  pages={5},
  year={2014},
  publisher={Frontiers Media SA}
}

@article{farhi2014quantum,
  title={A quantum approximate optimization algorithm},
  author={Farhi, Edward and Goldstone, Jeffrey and Gutmann, Sam},
  journal={arXiv preprint arXiv:1411.4028},
  year={2014}
}

@book{hadfield2018quantum,
  title={Quantum algorithms for scientific computing and approximate optimization},
  author={Hadfield, Stuart Andrew},
  year={2018},
  publisher={Columbia University}
}

@article{chancellor2019domain,
  title={Domain wall encoding of discrete variables for quantum annealing and {QAOA}},
  author={Chancellor, Nicholas},
  journal={Quantum Science and Technology},
  volume={4},
  number={4},
  pages={045004},
  year={2019},
  publisher={IOP Publishing}
}

@article{barenco1995elementary,
  title={Elementary gates for quantum computation},
  author={Barenco, Adriano and Bennett, Charles H and Cleve, Richard and DiVincenzo, David P and Margolus, Norman and Shor, Peter and Sleator, Tycho and Smolin, John A and Weinfurter, Harald},
  journal={Physical Review A},
  volume={52},
  number={5},
  pages={3457},
  year={1995},
  publisher={APS}
}

@article{abbas2024challenges,
  title={Challenges and opportunities in quantum optimization},
  author={Abbas, Amira and Ambainis, Andris and Augustino, Brandon and B{\"a}rtschi, Andreas and Buhrman, Harry and Coffrin, Carleton and Cortiana, Giorgio and Dunjko, Vedran and Egger, Daniel J and Elmegreen, Bruce G and others},
  journal={Nature Reviews Physics},
  pages={1--18},
  year={2024},
  publisher={Nature Publishing Group}
}

@article{nguyen2024empowering,
  title={Empowering a qudit-based quantum processor by traversing the dual bosonic ladder},
  author={Nguyen, Long B and Goss, Noah and Siva, Karthik and Kim, Yosep and Younis, Ed and Qing, Bingcheng and Hashim, Akel and Santiago, David I and Siddiqi, Irfan},
  journal={Nature Communications},
  volume={15},
  number={1},
  pages={7117},
  year={2024},
  publisher={Nature Publishing Group UK London}
}

@article{chi2022programmable,
  title={A programmable qudit-based quantum processor},
  author={Chi, Yulin and Huang, Jieshan and Zhang, Zhanchuan and Mao, Jun and Zhou, Zinan and Chen, Xiaojiong and Zhai, Chonghao and Bao, Jueming and Dai, Tianxiang and Yuan, Huihong and others},
  journal={Nature Communications},
  volume={13},
  number={1},
  pages={1166},
  year={2022},
  publisher={Nature Publishing Group UK London}
}

@article{roy2023two,
  title={Two-qutrit quantum algorithms on a programmable superconducting processor},
  author={Roy, Tanay and Li, Ziqian and Kapit, Eliot and Schuster, DavidI},
  journal={Physical Review Applied},
  volume={19},
  number={6},
  pages={064024},
  year={2023},
  publisher={APS}
}

@article{denys2023,
  title={The $2 T $-qutrit, a two-mode bosonic qutrit},
  author={Denys, Aur{\'e}lie and Leverrier, Anthony},
  journal={Quantum},
  volume={7},
  pages={1032},
  year={2023},
  publisher={Verein zur F{\"o}rderung des Open Access Publizierens in den Quantenwissenschaften}
}

@article{bornman2025benchmarking,
  title={Benchmarking the performance of a high-Q cavity qudit using random unitaries},
  author={Bornman, Nicholas and Roy, Tanay and Job, Joshua A and Anand, Namit and Perdue, Gabriel N and Zorzetti, Silvia and Alam, M Sohaib},
  journal={Quantum Science and Technology},
  volume={10},
  number={2},
  pages={025062},
  year={2025},
  publisher={IOP Publishing}
}

@article{su2021construction,
  title={Construction of a qudit using {S}chr{\"o}dinger cat states and generation of hybrid entanglement between a discrete-variable qudit and a continuous-variable qudit},
  author={Su, Qi-Ping and Liu, Tong and Zhang, Yu and Yang, Chui-Ping},
  journal={Physical Review A},
  volume={104},
  number={3},
  pages={032412},
  year={2021},
  publisher={APS}
}

@article{kim2025ultracoherent,
  title={Ultracoherent superconducting cavity-based multiqudit platform with error-resilient control},
  author={Kim, Taeyoon and Roy, Tanay and You, Xinyuan and Li, Andy CY and Lamm, Henry and Pronitchev, Oleg and Bal, Mustafa and Garattoni, Sabrina and Crisa, Francesco and Bafia, Daniel and others},
  journal={arXiv preprint arXiv:2506.03286},
  year={2025}
}

@book{ausiello2012complexity,
  title={Complexity and approximation: Combinatorial optimization problems and their approximability properties},
  author={Ausiello, Giorgio and Crescenzi, Pierluigi and Gambosi, Giorgio and Kann, Viggo and Marchetti-Spaccamela, Alberto and Protasi, Marco},
  year={2012},
  publisher={Springer Science \& Business Media}
}

@article{wang2020qudits,
  title={Qudits and high-dimensional quantum computing},
  author={Wang, Yuchen and Hu, Zixuan and Sanders, Barry C and Kais, Sabre},
  journal={Frontiers in Physics},
  volume={8},
  pages={589504},
  year={2020},
  publisher={Frontiers Media SA}
}

@article{deller2023quantum,
  title={Quantum approximate optimization algorithm for qudit systems},
  author={Deller, Yannick and Schmitt, Sebastian and Lewenstein, Maciej and Lenk, Steve and Federer, Marika and Jendrzejewski, Fred and Hauke, Philipp and Kasper, Valentin},
  journal={Physical Review A},
  volume={107},
  number={6},
  pages={062410},
  year={2023},
  publisher={APS}
}

@article{shaydulin2024evidence,
  title={Evidence of scaling advantage for the quantum approximate optimization algorithm on a classically intractable problem},
  author={Shaydulin, Ruslan and Li, Changhao and Chakrabarti, Shouvanik and DeCross, Matthew and Herman, Dylan and Kumar, Niraj and Larson, Jeffrey and Lykov, Danylo and Minssen, Pierre and Sun, Yue and others},
  journal={Science Advances},
  volume={10},
  number={22},
  pages={eadm6761},
  year={2024},
  publisher={American Association for the Advancement of Science}
}

@article{liu2024hybrid,
  title={Hybrid oscillator-qubit quantum processors: Instruction set architectures, abstract machine models, and applications},
  author={Liu, Yuan and Singh, Shraddha and Smith, Kevin C and Crane, Eleanor and Martyn, John M and Eickbusch, Alec and Schuckert, Alexander and Li, Richard D and Sinanan-Singh, Jasmine and Soley, Micheline B and others},
  journal={arXiv preprint arXiv:2407.10381},
  year={2024}
}

@article{di2021improving,
  title={Improving {H}amiltonian encodings with the {G}ray code},
  author={Di Matteo, Olivia and McCoy, Anna and Gysbers, Peter and Miyagi, Takayuki and Woloshyn, RM and Navr{\'a}til, Petr},
  journal={Physical Review A},
  volume={103},
  number={4},
  pages={042405},
  year={2021},
  publisher={APS}
}

@article{fuchs2021efficient,
  title={Efficient encoding of the weighted max k-cut on a quantum computer using {QAOA}},
  author={Fuchs, Franz G and Kolden, Herman {\O}ie and Aase, Niels Henrik and Sartor, Giorgio},
  journal={SN Computer Science},
  volume={2},
  number={2},
  pages={89},
  year={2021},
  publisher={Springer}
}

@article{heeres2015cavity,
  title={Cavity state manipulation using photon-number selective phase gates},
  author={Heeres, Reinier W and Vlastakis, Brian and Holland, Eric and Krastanov, Stefan and Albert, Victor V and Frunzio, Luigi and Jiang, Liang and Schoelkopf, Robert J},
  journal={Physical Review Letters},
  volume={115},
  number={13},
  pages={137002},
  year={2015},
  publisher={APS}
}

@article{krastanov2015universal,
  title={Universal control of an oscillator with dispersive coupling to a qubit},
  author={Krastanov, Stefan and Albert, Victor V and Shen, Chao and Zou, Chang-Ling and Heeres, Reinier W and Vlastakis, Brian and Schoelkopf, Robert J and Jiang, Liang},
  journal={Physical Review A},
  volume={92},
  number={4},
  pages={040303},
  year={2015},
  publisher={APS}
}

@article{eickbusch2022fast,
  title={Fast universal control of an oscillator with weak dispersive coupling to a qubit},
  author={Eickbusch, Alec and Sivak, Volodymyr and Ding, Andy Z and Elder, Salvatore S and Jha, Shantanu R and Venkatraman, Jayameenakshi and Royer, Baptiste and Girvin, Steven M and Schoelkopf, Robert J and Devoret, Michel H},
  journal={Nature Physics},
  volume={18},
  number={12},
  pages={1464--1469},
  year={2022},
  publisher={Nature Publishing Group UK London}
}

@article{kurkccuoglu2024qudit,
  title={Qudit Gate Decomposition Dependence for Lattice Gauge Theories},
  author={K{\"u}rk{\c{c}}{\"u}oglu, Doga Murat and Lamm, Henry and Maestri, Andrea},
  journal={arXiv preprint arXiv:2410.16414},
  year={2024}
}

@inproceedings{Gokhale2019QutritCircuits,
  author    = {Gokhale, Pranav and Baker, Jonathan M. and Duckering, Charlie and Brown, Neal and Brown, Kenneth R. and Chong, Frederic T.},
  title     = {Asymptotic Improvements to Quantum Circuits via Qutrits},
  booktitle = {Proceedings of the 46th International Symposium on Computer Architecture (ISCA)},
  pages     = {554--566},
  year      = {2019},
  publisher = {ACM},
  doi       = {10.1145/3307650.3322253}
}

@article{Blok2021QutritScrambling,
  author  = {Blok, M. S. and Ramasesh, V. V. and Schuster, D. I. and Siddiqi, I.},
  title   = {Quantum Information Scrambling on a Superconducting Qutrit Processor},
  journal = {Physical Review X},
  volume  = {11},
  pages   = {021010},
  year    = {2021},
  doi     = {10.1103/PhysRevX.11.021010}
}

@article{Ringbauer2022UniversalQudit,
  author  = {Ringbauer, Martin and Meth, Matthias and Postler, Lukas and Stricker, Roman and Pogorelov, Ilya and Lanyon, Benjamin P. and Blatt, Rainer and Monz, Thomas},
  title   = {Universal Qudit Quantum Computation with Trapped Ions},
  journal = {Nature Physics},
  volume  = {18},
  pages   = {1053--1057},
  year    = {2022},
  doi     = {10.1038/s41567-022-01640-6}
}

@article{tripathi2025qudit,
  title={Qudit dynamical decoupling on a superconducting quantum processor},
  author={Tripathi, Vinay and Goss, Noah and Vezvaee, Arian and Nguyen, Long B and Siddiqi, Irfan and Lidar, Daniel A},
  journal={Physical Review Letters},
  volume={134},
  number={5},
  pages={050601},
  year={2025},
  publisher={APS}
}

@article{Knill1996NonBinary,
  title     = {Non-binary Unitary Error Bases and Quantum Codes},
  author    = {Knill, Emanuel},
  journal   = {arXiv preprint quant-ph/9608048},
  year      = {1996}
}

@article{Gottesman1999FaultTolerant,
  title     = {Fault-Tolerant Quantum Computation with Higher-Dimensional Systems},
  author    = {Gottesman, Daniel},
  journal   = {Chaos, Solitons \& Fractals},
  volume    = {10},
  number    = {10},
  pages     = {1749--1758},
  year      = {1999},
  doi       = {10.1016/S0960-0779(99)00102-2}
}

@article{Ashikhmin2001Nonbinary,
  title     = {Nonbinary Quantum Stabilizer Codes},
  author    = {Ashikhmin, Alexei and Knill, Emanuel},
  journal   = {IEEE Transactions on Information Theory},
  volume    = {47},
  number    = {7},
  pages     = {3065--3072},
  year      = {2001},
  doi       = {10.1109/18.959288}
}

@article{Hostens2005Stabilizer,
  title={Stabilizer states and {C}lifford operations for systems of arbitrary dimensions and modular arithmetic},
  author={Hostens, Erik and Dehaene, Jeroen and De Moor, Bart},
  journal={Physical Review A},
  volume={71},
  number={4},
  pages={042315},
  year={2005},
  publisher={APS}
}

@article{bullock2007qudit,
  title={Qudit surface codes and gauge theory with finite cyclic groups},
  author={Bullock, Stephen S and Brennen, Gavin K},
  journal={Journal of Physics A: Mathematical and Theoretical},
  volume={40},
  number={13},
  pages={3481},
  year={2007},
  publisher={IOP Publishing}
}

@article{dutta2025noise,
  title={Noise-adapted qudit codes for amplitude-damping noise},
  author={Dutta, Sourav and Biswas, Debjyoti and Mandayam, Prabha},
  journal={Physical Review A},
  volume={111},
  number={3},
  pages={032438},
  year={2025},
  publisher={APS}
}

@article{grassl2018quantum,
  title={Quantum error-correcting codes for qudit amplitude damping},
  author={Grassl, Markus and Kong, Linghang and Wei, Zhaohui and Yin, Zhang-Qi and Zeng, Bei},
  journal={IEEE Transactions on Information Theory},
  volume={64},
  number={6},
  pages={4674--4685},
  year={2018},
  publisher={IEEE}
}

@article{chuang1997bosonic,
  title={Bosonic quantum codes for amplitude damping},
  author={Chuang, Isaac L and Leung, Debbie W and Yamamoto, Yoshihisa},
  journal={Physical Review A},
  volume={56},
  number={2},
  pages={1114},
  year={1997},
  publisher={APS}
}

@article{ozguler2022numerical,
  title={Numerical gate synthesis for quantum heuristics on bosonic quantum processors},
  author={{\"O}zg{\"u}ler, A Bar{\i}{\c{s}} and Venturelli, Davide},
  journal={Frontiers in Physics},
  volume={10},
  pages={900612},
  year={2022},
  publisher={Frontiers Media SA}
}

@article{bravyi2022hybrid,
  title={Hybrid quantum-classical algorithms for approximate graph coloring},
  author={Bravyi, Sergey and Kliesch, Alexander and Koenig, Robert and Tang, Eugene},
  journal={Quantum},
  volume={6},
  pages={678},
  year={2022},
  publisher={Verein zur F{\"o}rderung des Open Access Publizierens in den Quantenwissenschaften}
}

@article{apte2026quantum,
  title={Quantum Approximate Optimization of Integer Graph Problems and Surpassing Semidefinite Programming for Max-k-Cut},
  author={Apte, Anuj and Boulebnane, Sami and Jin, Yuwei and Omanakuttan, Sivaprasad and Perlin, Michael A and Shaydulin, Ruslan},
  journal={arXiv preprint arXiv:2602.05956},
  year={2026}
}

@inproceedings{maciejewski2024multilevel,
  author = {Maciejewski, Filip B and Bach, Bao G and Dupont, Maxime and Lott, P Aaron and Sundar, Bhuvanesh and Neira, David E Bernal and Safro, Ilya and Venturelli, Davide},
  booktitle = {2024 IEEE High Performance Extreme Computing Conference (HPEC)},
  doi = {10.1109/hpec62836.2024.10938438},
  organization = {IEEE},
  pages = {1--10},
  title = {{A} multilevel approach for solving large-scale {QUBO} problems with noisy hybrid quantum approximate optimization},
  year = {2024}
}

@article{marshall2026quantum,
  title={Quantum-enhanced {M}arkov Chain {M}onte {C}arlo for Combinatorial Optimization},
  author={Marshall, Kate V and Egger, Daniel J and Garn, Michael and Schiavello, Francesca and Brandhofer, Sebastian and Zoufal, Christa and Woerner, Stefan},
  journal={arXiv preprint arXiv:2602.06171},
  year={2026}
}

@article{diez2023quantum,
  title={Quantum approximate optimization algorithm pseudo-{B}oltzmann states},
  author={D{\'\i}ez-Valle, Pablo and Porras, Diego and Garc{\'\i}a-Ripoll, Juan Jos{\'e}},
  journal={Physical Review Letters},
  volume={130},
  number={5},
  pages={050601},
  year={2023},
  publisher={APS}
}

@article{lotshaw2023approximate,
  title={Approximate {B}oltzmann distributions in quantum approximate optimization},
  author={Lotshaw, Phillip C and Siopsis, George and Ostrowski, James and Herrman, Rebekah and Alam, Rizwanul and Powers, Sarah and Humble, Travis S},
  journal={Physical Review A},
  volume={108},
  number={4},
  pages={042411},
  year={2023},
  publisher={APS}
}

@article{awschalom2025challenges,
  title={Challenges and opportunities for quantum information hardware},
  author={Awschalom, David D and Bernien, Hannes and Hanson, Ronald and Oliver, William D and Vu{\v{c}}kovi{\'c}, Jelena},
  journal={Science},
  volume={390},
  number={6777},
  pages={1004--1010},
  year={2025},
  publisher={American Association for the Advancement of Science}
}

@article{wang2020xy,
  title={{XY} mixers: Analytical and numerical results for the quantum alternating operator ansatz},
  author={Wang, Zhihui and Rubin, Nicholas C and Dominy, Jason M and Rieffel, Eleanor G},
  journal={Physical Review A},
  volume={101},
  number={1},
  pages={012320},
  year={2020},
  publisher={APS}
}

@article{suri2024uniformly,
  title={Uniformly decaying subspaces for error-mitigated quantum computation},
  author={Suri, Nishchay and Saied, Jason and Venturelli, Davide},
  journal={Physical Review A},
  volume={110},
  number={4},
  pages={042621},
  year={2024},
  publisher={APS}
}

@article{dambal2025harnessing,
  title={Harnessing intrinsic noise for quantum simulation of open quantum systems},
  author={Dambal, Sameer and Sone, Akira and Zhang, Yu},
  journal={arXiv preprint arXiv:2510.21075},
  year={2025}
}

@article{newpaperunpublished,
  title={Quantum Approximate Optimization via Noise-Directed Adaptive Warm-Starting},
  author={Maciejewski, Filip and Hadfield, Stuart and Wallis, Oscar and  Pennington, George and Brandhofer, Sebastian and Woerner, Stefan and  Egger, Daniel J. and Venturelli, Davide},
  journal={arXiv preprint arXiv:2607.09368},
  year={2026}
}

@article{wood2018quantification,
  title={Quantification and characterization of leakage errors},
  author={Wood, Christopher J and Gambetta, Jay M},
  journal={Physical Review A},
  volume={97},
  number={3},
  pages={032306},
  year={2018},
  publisher={APS}
}

@article{li2025universal,
  title={Universal pulses for superconducting qudit ladder gates},
  author={Li, Boxi and C{\'a}rdenas-L{\'o}pez, FA and Lupascu, Adrian and Motzoi, Felix},
  journal={PRX Quantum},
  volume={6},
  number={3},
  pages={030357},
  year={2025},
  publisher={APS}
}

@article{lu2023high,
  title={High-fidelity parametric beamsplitting with a parity-protected converter},
  author={Lu, Yao and Maiti, Aniket and Garmon, John W O and Ganjam, Suhas and Liu, Yaxing and Cao, Jahan and Chapman, Benjamin J and Tsunoda, Takahiro and Eickbusch, Alec and Schoelkopf, Robert J and Devoret, Michel H},
  journal={Nature Communications},
  volume={14},
  pages={5767},
  year={2023},
  publisher={Nature Publishing Group UK London}
}

@article{tsunoda2023error,
  title={Error-detectable bosonic entangling gates with a noisy ancilla},
  author={Tsunoda, Takahiro and Teoh, James D and Kalfus, William D and Rosenblum, Serge and Eickbusch, Alec and Frattini, Nicholas E and Reinhold, Philip and Schoelkopf, Robert J and Devoret, Michel H and Jiang, Liang and Girvin, Steven M},
  journal={PRX Quantum},
  volume={4},
  number={2},
  pages={020354},
  year={2023},
  publisher={American Physical Society}
}

@article{huang2025fast,
  title={Fast sideband control of a weakly coupled multimode bosonic memory},
  author={Huang, Yuwei and DiNapoli, Stefano and Rockwood, Gregory and Vrajitoarea, Andrei and Boutin, Samuel and Ma, Wen-Long and Roy, Tanay and Chakram, Srivatsan},
  journal={arXiv preprint arXiv:2503.10623},
  year={2025}
}

@article{diringer2024conditional,
  title={Conditional-not displacement: Fast multioscillator control with a single qubit},
  author={Diringer, Asaf A and Blumenthal, Eli and Grinberg, Avishay and Jiang, Liang and Hacohen-Gourgy, Shay},
  journal={Physical Review X},
  volume={14},
  number={1},
  pages={011055},
  year={2024},
  publisher={American Physical Society}
}

@article{lapointemajor2024optimizing,
  title={Optimizing pulse shapes of an echoed conditional displacement gate in a superconducting bosonic system},
  author={Lapointe-Major, Maxime and others},
  journal={arXiv preprint arXiv:2408.05299},
  year={2024}
}

@article{dutta2025simulating,
  title={Simulating electronic structure on bosonic quantum computers},
  author={Dutta, Rishab and Cao, Ningyi and Bhattacharyya, Aniruddha and Mead, James and Pal, Anirban and others},
  journal={Journal of Chemical Theory and Computation},
  volume={21},
  pages={2281},
  year={2025},
  publisher={American Chemical Society}
}

@article{proctor2017ancilla,
  title={Ancilla-driven quantum computation for qudits and continuous variables},
  author={Proctor, Timothy and Giulian, Melissa and Korolkova, Natalia and Andersson, Erika and Kendon, Viv},
  journal={Physical Review A},
  volume={95},
  number={5},
  pages={052317},
  year={2017},
  publisher={APS}
}

@article{yang2025quantum,
  title={Quantum circuit synthesis with qudit phase gadget method},
  author={Yang, Shuai and Xu, Lihao and Tian, Guojing and Sun, Xiaoming},
  journal={arXiv preprint arXiv:2504.12710},
  year={2025}
}

@article{murairi2024highly,
  title={Highly-efficient quantum {F}ourier transformations for certain non-{A}belian groups},
  author={Murairi, Edison M and Alam, M Sohaib and Lamm, Henry and Hadfield, Stuart and Gustafson, Erik},
  journal={Physical Review D},
  volume={110},
  number={7},
  pages={074501},
  year={2024},
  publisher={APS}
}

@article{job2023efficient,
  title={Efficient, direct compilation of {SU(N)} operations into {SNAP} \& Displacement gates},
  author={Job, Joshua},
  journal={arXiv preprint arXiv:2307.11900},
  year={2023}
}

@article{ogunkoya2025investigating,
  title={Investigating parameter trainability in the {SNAP}-displacement protocol of a qudit system},
  author={Ogunkoya, Oluwadara and Morris, Kirsten and K{\"u}rk{\c{c}}{\"u}oglu, Doga Murat},
  journal={Physica Scripta},
  volume={100},
  number={7},
  pages={075109},
  year={2025},
  publisher={IOP Publishing}
}

@article{ozguler2024dynamics,
  title={Dynamics of qudit gates and effects of spectator modes on optimal control pulses},
  author={{\"O}zg{\"u}ler, A Bar{\i}{\c{s}} and Job, Joshua A},
  journal={Physical Review A},
  volume={109},
  number={5},
  pages={052404},
  year={2024},
  publisher={APS}
}

@inproceedings{reineri2023exploration,
  title={Exploration of superconducting multi-mode cavity architectures for quantum computing},
  author={Reineri, Alessandro and Zorzetti, Silvia and Roy, Tanay and You, Xinyuan},
  booktitle={2023 IEEE International Conference on Quantum Computing and Engineering (QCE)},
  volume={1},
  pages={1342--1348},
  year={2023},
  organization={IEEE}
}

@article{godwood2026fault,
  title={Fault-Tolerant Resource Comparison of Qudit and Qubit Encodings for Diagonal Quadratic Operators},
  author={Godwood, Samuel and K{\"u}rk{\c{c}}{\"u}o{\u{g}}lu, Do{\u{g}}a Murat and Perdue, Gabriel N and Maneyro, Marina and Roggero, Alessandro},
  journal={arXiv preprint arXiv:2604.26792},
  year={2026}
}

@article{lotshaw2026iterative,
  title={Iterative warm-start optimization with quantum imaginary time evolution},
  author={Lotshaw, Phillip C and Morris, Titus and Hadfield, Stuart and Bennink, Ryan},
  journal={arXiv preprint arXiv:2604.26047},
  year={2026}
}

@article{fuchs2025encodings,
  title={Encodings of the weighted {MAX k-CUT} problem on qubit systems},
  author={Fuchs, Franz G and Pariente Bassa, Ruben and Lien, Frida},
  journal={Frontiers in Quantum Science and Technology},
  volume={4},
  pages={1636042},
  year={2025},
  publisher={Frontiers Media SA}
}

@article{leipold2024imposing,
  title={Imposing constraints on driver {H}amiltonians and mixing operators: From theory to practical implementation},
  author={Leipold, Hannes and Spedalieri, Federico and Hadfield, Stuart and Rieffel, Eleanor G},
  journal={ACM Transactions on Quantum Computing},
  year={2026},
  publisher={ACM New York, NY}
}

@article{stollenwerk2025measurement,
  title={Measurement-driven Quantum Approximate Optimization},
  author={Stollenwerk, Tobias and Hadfield, Stuart},
  journal={arXiv preprint arXiv:2512.21046},
  year={2025}
}

\end{document}